# Free inductive $K$-semialgebras

Z. Ésik[*]

Dept. of Computer Science

University of Szeged

Hungary

W. Kuich[†]

Inst. for Discrete Mathematics and Geoometry

TU Vienna

Austria

**Abstract**

We consider rational power series over an alphabet $\Sigma$ with coefficients in a ordered commutative semiring $K$ and characterize them as the free ordered $K$-semialgebras in various classes of ordered $K$-semialgebras equipped with a star operation satisfying the least pre-fixed point rule and/or its dual. The results are generalizations of Kozen's axiomatization of regular languages.

## 1 Introduction

The equational theory of regular languages has been studied since the late 1950's, cf. [8, 24, 25, 26]. Axiomatizations of regular languages usually give rise to characterizations of the algebras of regular languages as the free objects in certain axiomatic classes of algebraic structures, and vice versa. For example, Kozen's axiomatization [20] describes regular languages over an alphabet $\Sigma$ as the free algebra over $\Sigma$ in the class of idempotent semirings equipped with a unary star operation subject to the fixed point identity $a^* = aa^* + 1$ and the least pre-fixed point rule and its dual:

$$ax + b \leq x \quad \Rightarrow \quad a^*b \leq x$$
$$xa + b \leq x \quad \Rightarrow \quad ba^* \leq x$$

Here, $\leq$ refers to the semilattice order, so that $a \leq b$ iff $a + b = b$. The dual of the fixed point identity, i.e., the equation $a^*a + 1 = a^*$ is a consequence of these axioms. Independently, and at about the same time, Krob proved in [22] that the algebras of regular languages are also free in the strictly larger class defined by the fixed point identity and just the least pre-fixed point rule. (Nevertheless, all the important models such as languages, binary relations, or continuous idempotent semirings satisfy both the least pre-fixed point rule and its dual.)

[*]Partially supported by the TÁMOP-4.2.1/B-09/1/KONV-2010-0005 program of National Development Agency of Hungary, the Austrian-Hungarian Action Foundation, grant 77öu9, the National Foundation of Hungary for Scientific Research, grant no. K 75249, and by grant T10003 from Reykjavik University's Development Fund.

[†]Partially supported by the Austrian-Hungarian Action Foundation, grant 77öu9.

In this paper we offer some generalizations of Kozen's and Krob's results. We consider rational power series [3, 9, 23, 27, 28] with coefficients in a commutative ordered semiring $K$ equipped with a star operation satisfying the above fixed point identity and either the first, or both versions of the least pre-fixed point rule. It is known that for any such semiring $K$ and for any alphabet $\Sigma$, the collection of rational power series over $\Sigma$ with coefficients in $K$, equipped with the usual sum, product, and star operations and the pointwise order, also satisfies these axioms, cf. [13]. We provide sufficient conditions under which these algebras of rational series, either equipped with the pointwise order, or by the sum order when $K$ is sum ordered, may be characterized as the free algebras in the class of ordered $K$-semialgebras satisfying the fixed point identity and one or both forms of the least pre-fixed point rule. Kozen's and Krob's results cover the case when $K$ is the Boolean semiring. The proofs of our results rely on extensions of the methods developed in [20] and [4].

## 2  Inductive semirings and inductive semialgebras

In this paper, we consider *semirings* $S = (S, +, \cdot, 0, 1)$ with an absorptive zero and a multiplicative identity 1, cf. [17, 18, 19, 23]. Morphisms of semirings preserve the sum and product operations and the constants 0 and 1. We call a semiring $S$ an *ordered semiring* if $S$ is equipped with a partial order relation $\leq$ which is preserved by the sum and product operations. A semiring $S$ is *positively ordered* if $0 \leq s$ holds for all $s \in S$. Morphisms of (positively) ordered semirings preserve the partial order. Examples of semirings include all rings, the semiring $\mathbb{N}$ of natural numbers, the Boolean semiring $\mathbb{B}$ on the set $\{0, 1\}$ whose sum and product operations are disjunction and conjunction, or more generally, every bounded distributive lattice, and the language semiring over an alphabet $\Sigma$, whose sum and product operations are set union and concatenation, respectively. The usual ordering turns $\mathbb{N}$ into a positively ordered semiring. Also, any bounded distributive lattice and any language semiring is a positively ordered semiring with the usual order relation.

In any semiring $S$, we may define the relation $\preceq$ by $s \preceq s'$ iff there exists some $r$ with $s + r = s'$. The relation $\preceq$ is always a preorder preserved by the operations. We call an ordered semiring $S$ a *sum ordered semiring* if its partial order is the relation $\preceq$. It is clear that a sum ordered semiring is positively ordered. If a semiring $S$ is positively ordered by the relation $\leq$ and if $s \preceq s'$ holds for some $s, s' \in S$, then also $s \leq s'$. Note also that any semiring morphism $S \to S'$ from a sum ordered semiring $S$ to a positively ordered semiring $S'$ automatically preserves the order relation. The semiring $\mathbb{N}$ is sum ordered as is any bounded distributive lattice and any language semiring.

An important special case arises by considering (additively) *idempotent* semirings satisfying $1 + 1 = 1$. Any idempotent semiring $S$ is positively ordered by the relation $\preceq$ which agrees with the semilattice order $\leq$ defined by $a \leq b$ iff $a + b = b$, for all $a, b \in S$. Moreover, this relation is the only one turning an idempotent semiring into a positively ordered semiring. Thus, an idempotent semiring is a positively ordered semiring in a unique way, and any morphism between idempotent semirings preserves the partial order.

We recall from [13] that an *inductive semiring*[1] is an ordered semiring $S$ equipped with a star operation $^* : S \to S$ subject to two axioms:

$$a^* = aa^* + 1 \tag{1}$$

$$ax + b \leq x \implies a^*b \leq x \tag{2}$$

---

[1] Inductive semirings were called inductive $^*$-semirings in [13].

for all $a, b, x \in S$. In a *symmetric inductive semiring* $S$, also

$$xa + b \leq x \implies ba^* \leq x \qquad (3)$$

holds for $a, b, x \in S$. The *fixed point identity* (1) may be replaced by the inequality $aa^* + 1 \leq a^*$. Clearly, every commutative inductive semiring is a symmetric inductive semiring. The axioms (2) and (3) are respectively called the *least pre-fixed point rule* and the *dual least pre-fixed point rule*. A morphism of (symmetric) inductive semirings is an ordered semiring morphism which *preserves the star operation*.

It is known that every inductive semiring $S$ is positively ordered and the star operation preserves the order: If $a \leq b$ in $S$, then $a^* \leq b^*$. Also, $a^*a + 1 = a^*$ holds for all $a \in S$.

The notion of symmetric inductive semiring generalizes the notion of *Kleene algebra* as defined in [20]. All "natural" inductive semirings are symmetric. The main examples of inductive semirings are the *continuous semirings* [17, 13]. There are various definitions of a continuous semirings in the literature. Here we adopt the following one. We say that a semiring $S$ is continuous if it is positively ordered, any nonempty countable directed set $X \subseteq S$ has a supremum, and the operations preserve suprema of such directed sets. Morphisms of continuous semirings preserve suprema of (nonempty) countable directed sets and hence the star operation defined by $a^* = \sup_n \sum_{i=0}^n a^i$. The fact that every continuous semiring $S$ is a symmetric inductive semiring is immediate, since for any $a, b \in S$, the endofunctions $S \to S$, $x \mapsto ax + b$ and $x \mapsto xa + b$ have as least (pre-)fixed points

$$\sup_n(b + ab + \cdots + a^n b) = (\sup_n(1 + a + \cdots + a^n))b = a^*b$$
$$\sup_n(b + ba + \cdots + ba^n) = b(\sup_n(1 + a + \cdots + a^n)) = ba^*.$$

Clearly, every finite positively ordered semiring is continuous. The class of continuous semirings includes all *quantales* and all *completely distributive complete lattices* and thus all language semirings and all finite distributive lattices. The existence of an inductive semiring (in fact "left handed Kleene algebra") that is not symmetric was pointed out in [21].

Since a finite positively ordered semiring is continuous, it is a symmetric inductive semiring. More generally, every locally finite[2] positively ordered semiring $S$ is a symmetric inductive semiring, since the sequences $(\sum_{i=0}^n a^i b)_n$ and $(\sum_{i=0}^n ba^i)_n$ are eventually finite for all $a, b \in S$. In particular, every bounded distributive lattice is a symmetric inductive semiring.

**Remark 2.1** *It is known, cf. [13], that a sum ordered semiring $S$ equipped with a star operation is an inductive semiring iff it satisfies the fixed point identity (1) and the axiom*

$$ax + b = x \implies a^*b \leq x \qquad (4)$$

*A similar fact holds for sum ordered symmetric inductive semirings, cf. [13].*

**Remark 2.2** *An* iteration semiring *is a semiring $S$ equipped with a star operation subject to an infinite collection of equational axioms, cf. [11, 10], consisting of the* sum star *and* product star *identities*

$$(a + b)^* = (a^*b)^*a^* \qquad (5)$$
$$(ab)^* = 1 + a(ba)^*b \qquad (6)$$

*and Conway's* group identities *[8, 11] associated with the finite (simple) groups. It follows that the identities $a^* = aa^* + 1$ and $a^* = a^*a + 1$ hold in all iteration semirings. It is known that every inductive semiring is an iteration semiring, cf. [13]. Below we will use this fact to derive certain properties of inductive semirings from corresponding properties of iteration semirings.*

---

[2]A semiring $S$ is locally finite if its finitely generated subsemirings are finite.

If $S$ is a (positively) ordered semiring, then equipped with the pointwise order, so is each *matrix semiring* $S^{n\times n}$ and each *formal series* semiring $S\langle\!\langle\Sigma^*\rangle\!\rangle$, cf. [3, 9, 18, 23], where $\Sigma$ is any alphabet. Here, $\Sigma^*$ denotes the free monoid of all words over $\Sigma$ including the empty word $\epsilon$. If $S$ is sum ordered, then so are $S^{n\times n}$ and $S\langle\!\langle\Sigma^*\rangle\!\rangle$, and the sum order agrees with the pointwise order. It was proved in [13] that if $S$ is a (symmetric) inductive semiring, then $S^{n\times n}$ and $S\langle\!\langle\Sigma^*\rangle\!\rangle$ are also (symmetric) inductive semirings. The star operation in $S^{n\times n}$ is determined by the matrix formula. Let $M \in S^{n\times n}$. When $n = 1$, $M = (a)$, for some $a \in S$, and we define $M^* = (a^*)$. Suppose now that $M = \begin{pmatrix} X & Y \\ U & V \end{pmatrix}$ where $X \in S^{k\times k}$, $Y \in S^{k\times 1}$, $U \in S^{1\times k}$ and $V \in S^{1\times 1}$, and suppose that we have already defined the star of any matrix over $S$ of size $m \times m$, where $m < n$. Then we define

$$\begin{pmatrix} X & Y \\ U & V \end{pmatrix}^* = \begin{pmatrix} \alpha & \beta \\ \gamma & \delta \end{pmatrix} \tag{7}$$

where

$$\begin{aligned} \alpha &= (X + YV^*U)^* & \beta &= \alpha YV^* \\ \gamma &= \delta U X^* & \delta &= (V + UX^*Y)^*. \end{aligned}$$

The star operation in $S\langle\!\langle\Sigma^*\rangle\!\rangle$ is given by $(r^*, \epsilon) = (r, \epsilon)^*$ and

$$(r^*, w) = \sum_{w_1\ldots w_n = w,\ w_i \neq \epsilon} (r, \epsilon)^*(r, w_1)(r, \epsilon)^* \cdots (r, w_n)(r, \epsilon)^*,$$

for all $w \in \Sigma^*$, $w \neq \epsilon$, and for all $r \in S\langle\!\langle\Sigma^*\rangle\!\rangle$, cf. [13]. It is known that this star operation is the unique extension of the star operation in $S$ that turns $S\langle\!\langle\Sigma^*\rangle\!\rangle$ into an iteration semiring, in fact and inductive semiring, see [4].

## 2.1 Functorial star

**Definition 2.3** *Suppose that $S$ is an inductive semiring and $\mathcal{C}$ is a class of matrices in $S^{m\times n}$, $m, n \geq 1$. We say that $S$ has a functorial star with respect to $\mathcal{C}$ if for all $A \in S^{m\times m}$, $B \in S^{n\times n}$ and $C \in S^{m\times n}$, if $AC = CB$ and $C \in \mathcal{C}$, then $A^*C = CB^*$. We say that $S$ has a* strong functorial star *if it has a functorial star with respect to the class of all matrices over $S$ of any dimension.*

For later use we prove:

**Lemma 2.4** *Any symmetric inductive semiring $S$ has a strong functorial star.*

*Proof.* Suppose that $S$ is a symmetric inductive semiring and $A \in S^{m\times m}$, $B \in S^{n\times n}$ and $C \in S^{m\times n}$ with $AC = CB$. Then

$$\begin{aligned} A(CB^*) + C &= CBB^* + C \\ &= CB^* \end{aligned}$$

and thus $A^*C \leq CB^*$. Symmetrically,

$$\begin{aligned} (A^*C)B + C &= A^*AC + C \\ &= A^*C \end{aligned}$$

so that $CB^* \leq A^*C$. □

We call a 0-1 matrix over a semiring *functional* if each row contains exactly one occurrence of 1. A *dual functional matrix* is the transpose of a functional matrix. The diagonal entries of a diagonal invertible (square) matrix have a multiplicative inverse. For inductive semirings that are not necessarily symmetric, we have the following facts.

**Lemma 2.5** *Suppose that $S$ is an inductive semiring. Then $S$ has a functorial star with respect to all invertible diagonal matrices.*

*Proof.* This is proved in [15] for all iteration semirings. But any inductive semiring is an iteration semiring. □

Following [5], we call a semiring $S$ *atomistic* if for all $a_1, \ldots, a_m, b_1, \ldots, b_n \in S$, where $m, n \geq 1$, if $a_1 + \cdots + a_m = b_1 + \cdots + b_n$, then there exist $c_1, \ldots, c_k \in S$ and partitions $\{I_i : i = 1, \ldots, m\}$ and $\{J_j : j = 1, \ldots, n\}$ of the set $\{1, \ldots, k\}$ (where some sets of the partition may be empty) such that

$$a_i = \sum_{k \in I_i} c_k$$
$$b_j = \sum_{k \in J_j} c_k$$

for all $i = 1, \ldots, m$ and $j = 1, \ldots, n$.

The following fact is taken from [4, 5]:

**Lemma 2.6** *Suppose that $S$ is an atomistic iteration semiring, or an atomistic inductive semiring. Then $S$ has a functorial star with respect to all functional and dual functional matrices.*

## 2.2 Inductive semialgebras

A semiring is called *commutative* if its multiplication operation is commutative. When $K$ is a commutative semiring, a $K$-*semialgebra* is a semiring $A = (A, +, \cdot, 0, e)$ (where the multiplicative identity is usually written $e$) equipped with a (unitary, zero preserving) left $K$-action turning $A$ into a $K$-semimodule, such that

$$(ka)b = k(ab) = a(kb)$$

holds for all $a, b \in A$ and $k \in K$. A morphism of $K$-semialgebras is a semiring morphism preserving the action.

For the rest of this section, suppose that $K$ is a commutative semiring.

For example, $K^{n \times n}$ and $K\langle\!\langle \Sigma^* \rangle\!\rangle$ equipped with the pointwise $K$-action are $K$-semialgebras for all $n \geq 1$ and for all alphabets $\Sigma$. Any semiring is naturally an $\mathbb{N}$-semialgebra, and any idempotent semiring is a $\mathbb{B}$-semialgebra.

When $K$ is (positively) ordered, a (positively) ordered $K$-semialgebra is a $K$-semialgebra which is a (positively) ordered semiring such that the $K$-action preserves the ordering in both arguments. A morphism of (positively) ordered $K$-semialgebras also preserves the partial order.

It is clear that every commutative inductive semiring is a symmetric inductive semiring. Suppose that $K$ is a commutative inductive semiring. Then a (symmetric) inductive $K$-semialgebra is a positively ordered $K$-semialgebra which is a (symmetric) inductive semiring. Moreover, the star operation in $K$ is compatible with the star operation of $A$:

$$(ke)^* = k^*e \qquad (8)$$

for all $k \in K$. Morphisms of (symmetric) inductive $K$-semialgebras are ordered $K$-semialgebra morphisms that are morphisms of (symmetric) inductive semirings.

Examples of (symmetric) inductive $K$-semialgebras include the $K$-semialgebras $K\langle\!\langle \Sigma^* \rangle\!\rangle$ where $\Sigma$ is any alphabet and $K$ is a commutative inductive semiring. In particular, when $K$ is $\mathbb{B}$, then $K\langle\!\langle \Sigma^* \rangle\!\rangle$ is essentially the language semiring over $\Sigma$. The partial order is the pointwise order. When $K$ is a commutative continuous semiring, then $K\langle\!\langle \Sigma^* \rangle\!\rangle$, equipped with the pointwise partial order, is a symmetric inductive $K$-semialgebra.

## 3 Automata

In this section, we define automata in inductive $K$-semialgebras, where $K$ is a fixed commutative inductive semiring, and state a variant of the Kleene theorem from [4, 13].

Let $K$ denote a commutative inductive semiring.

Suppose that $A$ is an inductive $K$-semialgebra and $\Sigma \subseteq A$. Let $K\Sigma$ denote the $K$-semimodule generated by $\Sigma$ in $A$, i.e., the collection of all finite linear combinations of elements of $\Sigma$ with coefficients in $K$. A *(K-)automaton over $\Sigma$ in $A$* is a triplet

$$\mathcal{A} = (\alpha, M, \beta)$$

where $\alpha \in K^{1 \times n}$, $M \in (K\Sigma)^{n \times n}$, $\beta \in K^{n \times 1}$ are the *initial vector*, the *transition matrix* and the *final vector* of $\mathcal{A}$, respectively. The integer $n \geq 1$ is called the *dimension* of $\mathcal{A}$. The *behavior* of $\mathcal{A}$ is

$$|\mathcal{A}| := \alpha M^* \beta \in A.$$

Two automata are called *equivalent* if they have the same behavior.

**Definition 3.1** *Suppose that $A$ is an inductive $K$-semialgebra and $\Sigma \subseteq A$. We say that $a \in A$ is $K$-recognizable over $\Sigma$ in $A$ if there is a $K$-automaton $\mathcal{A}$ over $\Sigma$ in $A$ with $|\mathcal{A}| = a$. We let $\mathrm{Rec}_A(\Sigma)$ denote the set of all $K$-recognizable elements over $\Sigma$.*

**Definition 3.2** *Suppose that $A$ is an inductive $K$-semialgebra and $\Sigma \subseteq A$. We say that $a \in A$ is $K$-rational over $\Sigma$ in $A$ if $a$ can be generated from the elements of $\Sigma$ and $0$ by the operations $+$, $\cdot$, $^*$ and the $K$-action. We let $\mathrm{Rat}_A(\Sigma)$ denote the collection of all $K$-rational elements over $\Sigma$.*

The following Kleene-type result follows from Corollary 9.4.4 in [4]. See also [6, 13].

**Theorem 3.3** *For any inductive $K$-semialgebra $A$ and any $\Sigma \subseteq A$, $\mathrm{Rec}_A(\Sigma) = \mathrm{Rat}_A(\Sigma)$.*

For any alphabet $\Sigma$, let $K^{\mathrm{rec}}\langle\!\langle \Sigma^* \rangle\!\rangle := \mathrm{Rec}_{K\langle\!\langle \Sigma^* \rangle\!\rangle}(\Sigma)$ and $K^{\mathrm{rat}}\langle\!\langle \Sigma^* \rangle\!\rangle := \mathrm{Rat}_{K\langle\!\langle \Sigma^* \rangle\!\rangle}(\Sigma)$. A series in $K^{\mathrm{rec}}\langle\!\langle \Sigma^* \rangle\!\rangle$ is called *$K$-recognizable*, and a series in $K^{\mathrm{rat}}\langle\!\langle \Sigma^* \rangle\!\rangle$ is called *$K$-rational*, cf. [3, 27, 28].

**Proposition 3.4** *For any commutative inductive semiring $K$ and alphabet $\Sigma$, $K^{\mathrm{rat}}\langle\!\langle \Sigma^* \rangle\!\rangle$, equipped with the pointwise order, is a (symmetric) inductive $K$-semialgebra.*

*Proof.* By definition, $K^{\mathrm{rat}}\langle\!\langle\Sigma^*\rangle\!\rangle$ is closed under star. It was shown in [13] that equipped with the pointwise order, $K\langle\!\langle\Sigma^*\rangle\!\rangle$ is a (symmetric) inductive semiring. It follows easily that $K^{\mathrm{rat}}\langle\!\langle\Sigma^*\rangle\!\rangle$, equipped with the pointwise order, is a (symmetric) inductive $K$-semialgebra. □

As noted above, if $K$ is sum ordered, then the pointwise order on $K\langle\!\langle\Sigma^*\rangle\!\rangle$ agrees with the sum order relation. However, the pointwise order on $K^{\mathrm{rat}}\langle\!\langle\Sigma^*\rangle\!\rangle$ may not be the same as the sum order relation, since if $s, s' \in K^{\mathrm{rat}}\langle\!\langle\Sigma^*\rangle\!\rangle$ with $s \leq s'$ in the sum order, there may not exist a *rational* series $r$ with $s + r = s'$, cf. [3].

## 3.1 Simulations of automata

The notion of simulation was introduced in [4, 12] in order to relate equivalent automata. It can be traced back to Schützenberger's result on the minimization of weighted automata over fields, cf. [3]. Simulations over the Boolean semiring are implicit in [20]. Simulation is called "conjugacy" in [1, 2]. The notion of (functional) simulation is closely related to the notion of bisimulation, see [4].

In this section, we assume that $K$ is a commutative inductive semiring.

**Definition 3.5** *Suppose that $A$ is an inductive $K$-semialgebra and $\Sigma \subseteq A$. Let $\mathcal{A} = (\alpha, M, \beta)$ and $\mathcal{B} = (\gamma, N, \delta)$ be automata over $\Sigma$ in $A$ of dimension $m$ and $n$, respectively. We say that a matrix $X \in K^{m \times n}$ is a* simulation $\mathcal{A} \to \mathcal{B}$, *denoted $\mathcal{A} \to^X \mathcal{B}$, if*

$$\alpha X = \gamma, \quad MX = XN, \quad \beta = X\delta.$$

*A* functional simulation *(*dual functional simulation*) is a simulation by a functional (dual functional) matrix. A* diagonal simulation *is a simulation by a diagonal square matrix. An* invertible diagonal simulation *is a diagonal simulation by an invertible diagonal matrix.*

*We say that $\mathcal{A}$ and $\mathcal{B}$ are* simulation equivalent *if they can be connected by a finite chain of simulations, i.e., when there is a finite sequence of automata $\mathcal{C}_i$ together with matrices $X_i$ of appropriate size, for $i = 0, \ldots, k-1$, such that $\mathcal{C}_0 = \mathcal{A}$, $\mathcal{C}_k = \mathcal{B}$, and for each $0 \leq i < k$, either $\mathcal{C}_{i+1} \to^{X_i} \mathcal{C}_i$ or $\mathcal{C}_i \to^{X_i} \mathcal{C}_{i+1}$. Moreover, we say that $\mathcal{A}$ and $\mathcal{B}$ are* strongly simulation equivalent *if they can be connected by a finite chain of simulations consisting of functional, dual functional and invertible diagonal simulations.*

Note that if $\mathcal{A} \to^X \mathcal{B}$ and $\mathcal{B} \to^Y \mathcal{C}$, then $\mathcal{A} \to^{XY} \mathcal{C}$, so that we may require that in the above definition of simulation equivalence, the simulations are "alternating".

**Lemma 3.6** *Suppose that $A$ is a symmetric inductive $K$-semialgebra with $\Sigma \subseteq A$. Suppose that $\mathcal{A}$ and $\mathcal{B}$ are $K$-automata over $\Sigma$ in $A$ such that there is a simulation $\mathcal{A} \to \mathcal{B}$. Then $\mathcal{A}$ and $\mathcal{B}$ are equivalent.*

*Proof.* Let $\mathcal{A}$ and $\mathcal{B}$ be automata in $A$ over $\Sigma \subseteq A$ as in Definition 3.5, and suppose that $X \in K^{m \times n}$ is a simulation $\mathcal{A} \to \mathcal{B}$. Then, using Lemma 2.4,

$$\alpha M^* \beta = \alpha M^* X \delta = \alpha X N^* \delta = \gamma N^* \delta.$$

□

**Corollary 3.7** *Suppose that $A$ is a symmetric inductive $K$-semialgebra and that the $K$-automata $\mathcal{A}$ and $\mathcal{B}$ over $\Sigma \subseteq A$ are simulation equivalent. Then $\mathcal{A}$ and $\mathcal{B}$ are equivalent.*

For inductive $K$-semialgebras, we have weaker facts.

**Lemma 3.8** *Suppose that $A$ is an inductive $K$-semialgebra with $\Sigma \subseteq A$. Suppose that $\mathcal{A}$ and $\mathcal{B}$ are $K$-automata over $\Sigma$ in $A$ such that there is an invertible diagonal simulation $\mathcal{A} \to \mathcal{B}$. Then $\mathcal{A}$ and $\mathcal{B}$ are equivalent.*

*Proof.* Identical to the proof of Lemma 3.6 using Lemma 2.5. □

**Lemma 3.9** *Let $K$ be an atomistic commutative inductive semiring. Suppose that $A$ is an inductive $K$-semialgebra with $\Sigma \subseteq A$. Suppose that $\mathcal{A}$ and $\mathcal{B}$ are $K$-automata over $\Sigma$ in $A$ such that there is a functional or dual functional simulation $\mathcal{A} \to \mathcal{B}$. Then $\mathcal{A}$ and $\mathcal{B}$ are equivalent.*

*Proof.* Identical to the proof of Lemma 3.6 using Lemma 2.6. □

**Corollary 3.10** *Suppose that $A$ is an inductive $K$-semialgebra and that the $K$-automata $\mathcal{A}$ and $\mathcal{B}$ over $\Sigma \subseteq A$ are strongly simulation equivalent. Then $\mathcal{A}$ and $\mathcal{B}$ are equivalent.*

## 3.2 Proper and strongly proper semirings

Suppose that $K$ is a commutative inductive semiring. We call $K$ *proper* if for all alphabets $\Sigma$, whenever two $K$-automata over $\Sigma$ in $K\langle\!\langle \Sigma^* \rangle\!\rangle$ are equivalent, then they are simulation equivalent. Similarly, we call $K$ *strongly proper*, if whenever two $K$-automata over $\Sigma$ in $K\langle\!\langle \Sigma^* \rangle\!\rangle$ are equivalent then they are strongly simulation equivalent. The above definitions were extracted from [4, 12, 1] in [16]. Actually the definition given in [16] applies to all semirings, not just commutative inductive semirings.

A semiring $S$ is called *Noetherian* if every subsemimodule of a finitely generated $S$-semimodule is finitely generated. It is known that all (commutative) semirings whose finitely generated subsemirings are Noetherian or included in a Noetherian subsemiring are proper, cf. [16]. This includes all locally finite semirings and all commutative rings. In particular, any finite positively ordered commutative semiring and any bounded distributive lattice such as the Boolean semiring $\mathbb{B}$ is proper. There are examples of non-proper semirings, cf. [16]. It is known that if a semiring $S$ is "equisubstractive" and "additively generated by its multiplicatively invertible elements", then $S$ is strongly proper iff $S$ is proper, cf. [1]. There has not been any example of a semiring that is proper but not strongly proper.

# 4 Free inductive $K$-semialgebras

In this section, we assume that $K$ is a commutative inductive semiring. Our aim is to prove that under additional assumptions on $K$, for any alphabet $\Sigma$, $K^{\mathrm{rat}}\langle\!\langle \Sigma^* \rangle\!\rangle$, equipped either with the pointwise order or the sum order, is a free symmetric inductive, or even a free inductive $K$-semialgebra on $\Sigma$.

**Theorem 4.1** *Suppose that $\Sigma$ is an alphabet and $K$ is a commutative inductive semiring that is proper. Then $K^{\mathrm{rat}}\langle\!\langle \Sigma^* \rangle\!\rangle$ has the following property: Given any symmetric inductive $K$-semialgebra $A$ and function $h : \Sigma \to A$, there is a unique morphism of $K$-semialgebras $h^\sharp : K^{\mathrm{rat}}\langle\!\langle \Sigma^* \rangle\!\rangle \to A$ preserving the star operation which extends $h$.*

1. If $h^\sharp$ preserves the pointwise ordering of $K^{\mathrm{rat}}\langle\!\langle \Sigma^* \rangle\!\rangle$, for all $A$ and all functions $h$, then equipped with the pointwise order, $K^{\mathrm{rat}}\langle\!\langle \Sigma^* \rangle\!\rangle$ is a free symmetric inductive semiring on $\Sigma$.

2. If $K$ is sum ordered and $K^{\mathrm{rat}}\langle\!\langle \Sigma^* \rangle\!\rangle$, equipped with the sum order relation is a symmetric inductive semiring and hence a symmetric inductive $K$-semialgebra, then $K^{\mathrm{rat}}\langle\!\langle \Sigma^* \rangle\!\rangle$ is a free symmetric inductive $K$-semialgebra generated by $\Sigma$.

*Proof.* Suppose that $A$ is a symmetric inductive $K$-semialgebra and $h$ is a function $\Sigma \to A$. We show that $h$ can be extended in a unique way to a morphism $h^\sharp : K^{\mathrm{rat}}\langle\!\langle \Sigma^* \rangle\!\rangle \to A$ of inductive $K$-semialgebras preserving the star operation. First, we extend $h$ to linear combinations $a = k_1 a_1 + \cdots + k_m a_m$ where $k_i \in K$ and $a_i \in \Sigma$ for all $i = 1, \ldots, m$ by defining $ah := k_1(a_1 h) + \cdots + k_m(a_m h)$. Let $r \in K^{\mathrm{rat}}\langle\!\langle \Sigma^* \rangle\!\rangle$. Then $r$ is the behavior of some $K$-automaton $\mathcal{A} = (\alpha, M, \beta)$ over $\Sigma$ in $K^{\mathrm{rat}}\langle\!\langle \Sigma^* \rangle\!\rangle$. Let $\mathcal{A}h = (\alpha, Mh, \beta)$, where $Mh$ is defined pointwise. Clearly, $\mathcal{A}h$ is a $K$-automaton over $\Sigma h$ in $A$. Since the star operation on matrices is determined by the star operation in $K$, we are forced to define $rh^\sharp := |\mathcal{A}h| = \alpha(Mh)^*\beta$.

To complete the proof, we need show that $h^\sharp$ is a well-defined function and preserves the operations and constants and the partial order. To prove that $h^\sharp$ is well-defined, suppose that $\mathcal{A}$ and $\mathcal{B}$ are equivalent automata over $\Sigma$ in $K^{\mathrm{rat}}\langle\!\langle \Sigma^* \rangle\!\rangle$. Since $K$ is proper, this means that $\mathcal{A}$ and $\mathcal{B}$ are simulation equivalent so that there exists a finite sequence of simulations connecting them which involves the automata $\mathcal{A}_0 = \mathcal{A}$, $\mathcal{A}_1$, ..., $\mathcal{A}_{m-1}$, $\mathcal{A}_m = \mathcal{B}$ and the matrices $X_1, \ldots, X_m$, say. Now $\mathcal{A}h$ and $\mathcal{B}h$ are also simulation equivalent, since they can be connected by the chain of simulations involving the same matrices and the automata $\mathcal{A}_0 h = \mathcal{A}h$, $\mathcal{A}_1 h$, ..., $\mathcal{A}_{m-1}h$, $\mathcal{A}_m h = \mathcal{B}h$. Thus, by Lemma 3.6, $\mathcal{A}h$ and $\mathcal{B}h$ are equivalent. This proves that $h^\sharp$ is well-defined.

The fact that $h^\sharp$ preserves the constants $0$ and $e$ and the operations $+, \cdot$ and $^*$ can be proved by standard automata constructions described in the proof of Theorem 5.1 in [4], Chapter 9. Here, we only treat the star operation. Suppose that $r = |\mathcal{A}|$, where $\mathcal{A}$ is the automaton $(\alpha, M, \beta)$ over $\Sigma$ in $K\langle\!\langle \Sigma^* \rangle\!\rangle$. Then we define

$$\mathcal{A}^* := \left( \begin{pmatrix} \alpha & 1 \end{pmatrix}, \begin{pmatrix} (\beta\alpha)^* M & 0 \\ 0 & 0 \end{pmatrix}, \begin{pmatrix} (\beta\alpha)^*\beta \\ 1 \end{pmatrix} \right)$$

Then $\mathcal{A}^*$ is a $K$-automaton over $\Sigma$ in $K\langle\!\langle \Sigma^* \rangle\!\rangle$ and $\mathcal{A}^* h$ is an automaton over $\Sigma h$ in $A$. Also,

$$\begin{pmatrix} (\beta\alpha)^* M & 0 \\ 0 & 0 \end{pmatrix}^* = \begin{pmatrix} ((\beta\alpha)^* M)^* & 0 \\ 0 & e \end{pmatrix}$$

and thus

$$\begin{aligned}
|\mathcal{A}^*| &= \alpha((\beta\alpha)^* M)^*(\beta\alpha)^*\beta + e \\
&= \alpha(\beta\alpha + M)^*\beta + e \\
&= \alpha(M^*\beta\alpha)^* M^*\beta + e \\
&= \alpha(M^*\beta(\alpha M^*\beta)^*\alpha + e)M^*\beta + e \\
&= (\alpha M\beta(\alpha M^*\beta)^* + e)\alpha M^*\beta + e \\
&= (\alpha M^*\beta)^*\alpha M^*\beta + e \\
&= (\alpha M^*\beta)^* \\
&= |\mathcal{A}|^*,
\end{aligned}$$

since the sum star (5) and product star (6) identities hold for matrices, cf. [13]. A similar computation shows that $|\mathcal{A}^* h| = |\mathcal{A}h|^*$. Thus,

$$r^* h^\sharp = |\mathcal{A}^* h| = |\mathcal{A}h|^* = (rh^\sharp)^*.$$

Also, $h^\sharp$ preserves the $K$-action, since if $r = |\mathcal{A}|$ where $\mathcal{A} = (\alpha, M, \beta)$ as above, then $kr = |k\mathcal{A}|$ where $k\mathcal{A} := (k\alpha, M, \beta)$ and thus

$$(kr)h^\sharp = |(k\mathcal{A})h| = k|\mathcal{A}h| = k(rh^\sharp).$$

Since for every letter $a \in \Sigma$ it holds that

$$ah^\sharp = |(1, (a), 1)h| = |(1, (ah), 1)| = ah,$$

$h^\sharp$ extends $h$.

Let us now equip $K^{\text{rat}}\langle\!\langle \Sigma^* \rangle\!\rangle$ with the pointwise partial order. Then $K^{\text{rat}}\langle\!\langle \Sigma^* \rangle\!\rangle$ is a symmetric inductive $K$-semialgebra. Thus, if $h^\sharp$ preserves the partial order for all symmetric inductive $K$-semialgebras $A$ and for all functions $h : \Sigma \to A$, then $K^{\text{rat}}\langle\!\langle \Sigma^* \rangle\!\rangle$ is a free symmetric inductive $K$-semialgebra on $\Sigma$.

Last, suppose that $K$ is sum ordered and let us equip $K^{\text{rat}}\langle\!\langle \Sigma^* \rangle\!\rangle$ with the sum order. Moreover, suppose that $K^{\text{rat}}\langle\!\langle \Sigma^* \rangle\!\rangle$, equipped with this partial order, is a symmetric inductive semiring and thus a symmetric inductive $K$-semialgebra. Let $A$ be any symmetric inductive $K$-semialgebra and suppose that $h : \Sigma \to A$. Since $h^\sharp$ preserves sum, it follows that $h^\sharp$ preserves the partial order. Indeed, if $s \leq r$ in $K^{\text{rat}}\langle\!\langle \Sigma^* \rangle\!\rangle$ then there is some $t$ in $K^{\text{rat}}\langle\!\langle \Sigma^* \rangle\!\rangle$ with $s + t = r$. Thus, $sh^\sharp + th^\sharp = rh^\sharp$ which implies that $sh^\sharp \leq rh^\sharp$ in $A$. Hence, in this case, $K^{\text{rat}}\langle\!\langle \Sigma^* \rangle\!\rangle$, equipped with the sum order, is a free symmetric inductive $K$-semialgebra on $\Sigma$. □

One of the assumptions of the last claim of Theorem 4.1 is that $K^{\text{rat}}\langle\!\langle \Sigma^* \rangle\!\rangle$, equipped with the sum order, is itself a symmetric inductive $K$-semialgebra. Since equipped with the pointwise order, $K^{\text{rat}}\langle\!\langle \Sigma^* \rangle\!\rangle$ is always a symmetric inductive $K$-semialgebra, this condition is fulfilled whenever the pointwise order agrees with the sum order. This includes the case when $K$ is idempotent.

**Corollary 4.2** *Suppose that $K$ is a commutative idempotent inductive semiring that is proper. Then for every alphabet $\Sigma$, $K^{\text{rat}}\langle\!\langle \Sigma^* \rangle\!\rangle$ is a free symmetric idempotent inductive $K$-semialgebra generated by $\Sigma$.*

In order to state the next corollary, we need the following well-known fact (see e.g., [3], where the same fact is stated for finite semirings).

**Lemma 4.3** *Suppose that $S$ is an inductive semiring that is a locally finite semiring. Then for any alphabet $\Sigma$, a series $r \in S\langle\!\langle \Sigma^* \rangle\!\rangle$ is rational iff the set $\{(s, w) : w \in \Sigma^*\}$ is finite (i.e., $r$ is image finite), and for each $s \in S$, the set $\{w \in \Sigma^* : (r, w) = s\}$ is regular.*

**Corollary 4.4** *Suppose that $K$ is a commutative positively ordered semiring that is locally finite as a semiring and is thus an inductive semiring. Then for any alphabet $\Sigma$, $K^{\text{rat}}\langle\!\langle \Sigma^* \rangle\!\rangle$, equipped with the pointwise order, is a free symmetric inductive semiring on $\Sigma$.*

*Proof.* Since $K$ is locally finite, it is proper. Suppose that $A$ is symmetric inductive $K$-semialgebra. We have to show that for any alphabet $\Sigma$, if $h : K^{\text{rat}}\langle\!\langle \Sigma^* \rangle\!\rangle \to A$ is a morphism of $K$-semialgebras preserving star, where $K^{\text{rat}}\langle\!\langle \Sigma^* \rangle\!\rangle$ is equipped with the pointwise order, then $h$ preserves the partial order. To this end, let $r, r' \in K^{\text{rat}}\langle\!\langle \Sigma^* \rangle\!\rangle$ with $(r, w) \leq (r', w)$ for all $w \in \Sigma^*$. Since $r, r'$ are rational, by the above lemma and since regular languages are closed under the Boolean operations, there exist pairwise disjoint regular languages $R_1, \ldots, R_n$ and $k_1, \ldots, k_n, k'_1, \ldots, k'_n \in K$ with $k_1 \leq k'_1, \ldots, k_n \leq k'_n$ such that

$$r = \sum_{i=1}^{n} k_i s_i$$

$$r' = \sum_{i=1}^{n} k'_i s_i$$

where for each $i = 1, \ldots, n$, $s_i$ denotes the characteristic series of $R_i$, so that $(s_i, w) = 1$ if $w \in R_i$ and $(s, w) = 0$ otherwise, for all $w \in \Sigma^*$. Since $k_i \leq k'_i$, also $k_i(s_i h) \leq k'_i(s_i h)$, for all $i$. Thus,

$$rh = \sum_{i=1}^{n} k_i(s_i h) \leq \sum_{i=1}^{n} k'_i(s_i h) = r'h.$$

$\square$

Each series in $\mathbb{B}^{\mathrm{rat}}\langle\!\langle \Sigma^* \rangle\!\rangle$ may be identified with its support that is a regular language in $\Sigma^*$.

**Corollary 4.5** *(Kozen [20]) For any alphabet $\Sigma$, the semiring $\mathbb{B}^{\mathrm{rat}}\langle\!\langle \Sigma^* \rangle\!\rangle$ is the free Kleene algebra on $\Sigma$.*

*Proof.* This follows from either Corollary 4.2 or Corollary 4.4.  $\square$

Below we will make use of the following fact.

**Lemma 4.6** *Suppose that $S$ is a continuous semiring and $a_{ni} \in S$ for all $n, i \geq 0$. Moreover, suppose that the sequence $(a_{ni})_n$ is increasing for each fixed integer $i$. Then*

$$\sup_n \sum_i a_{ni} = \sum_i \sup_n a_{ni}.$$

*Proof.* First note that the left hand side exists since the sequence $(\sum_i a_{ni})_n$ is increasing. Then we calculate as follows:

$$\begin{aligned}
\sup_n \sum_i a_{ni} &= \sup_n \sup_m \sum_{i=0}^{m} a_{ni} \\
&= \sup_m \sup_n \sum_{i=0}^{m} a_{ni} \\
&= \sup_m \sum_{i=0}^{m} \sup_n a_{ni} \\
&= \sum_i \sup_n a_{ni}.
\end{aligned}$$

$\square$

**Proposition 4.7** *Suppose that $K$ is a sum ordered continuous commutative semiring satisfying $1 \leq k$ for all $k \in K$, $k \neq 0$. Then for any alphabet $\Sigma$, $K^{\mathrm{rat}}\langle\!\langle \Sigma^* \rangle\!\rangle$, equipped with the sum-order, is a symmetric inductive $K$-semialgebra.*

*Proof.* Suppose that $s, r \in K^{\mathrm{rat}}\langle\!\langle \Sigma^* \rangle\!\rangle$. We have to show that whenever $t \in K^{\mathrm{rat}}\langle\!\langle \Sigma^* \rangle\!\rangle$ is a solution of the equation $x = sx + r$, then there is a series $t' \in K^{\mathrm{rat}}\langle\!\langle \Sigma^* \rangle\!\rangle$ with $s^*r + t' = t$. (See Remark 2.1.) When $s$ is proper, i.e., $(s, \epsilon) = 0$, then $s = t$, since $s^*r$ is the unique solution of the equation $x = sx + r$, even in $K\langle\!\langle \Sigma^* \rangle\!\rangle$, cf. [3].

So suppose that $s$ is not proper and $t$ is rational with $t = st + r$. Since $s$ is not proper and since $k \geq 1$ for all $k \neq 0$, we have $s \geq 1$. Then for all $n$,

$$t = s^n t + s^{n-1} r + \cdots + sr + r.$$

Since $s \geq 1$, the sequence $(s^n t)_n$ is increasing and $\sup_n (s^n t)$ exists. Thus, by continuity of the operations,

$$\begin{aligned} s^* r + (\sup_n s^n) t &= \sup_n(s^n r + \cdots + sr + r) + \sup_n(s^n t) \\ &= \sup_n(s^n t + s^{n-1} r + \cdots + sr + r) \\ &= \sup_n t \\ &= t. \end{aligned}$$

To complete the proof, it suffices to show that $s^\omega = \sup_n s^n$ is rational.

Write $s = k + s_0$, where $k \in K$ and $s_0$ is proper. Then for all $n$,

$$s^n = k^n + \binom{n}{1} k^{n-1} s_0 + \binom{n}{2} k^{n-2} s_0^2 + \cdots + s_0^n.$$

Since $k \geq 1$, the sequence $(k^n)_n$ is increasing with supremum $k^\omega$, say. Moreover, $k' = \sup_n \sum_{i=1}^n k^\omega$ exists. Thus, by Lemma 4.6,

$$\begin{aligned} s^\omega &= \sup_n k^n + \sup_n \left( \binom{n}{1} k^{n-1} s_0 + \binom{n}{2} k^{n-2} s_0^2 + \cdots + s_0^n \right) \\ &= k^\omega + \sum_i \sup_n \binom{n}{i} k^{n-i} s_0^i \\ &= k^\omega + \sum_i k' s_0^i \\ &= k^\omega + k' s_0^+, \end{aligned}$$

proving that $s^\omega$ is rational. $\square$

**Corollary 4.8** *Suppose that $K$ is a sum ordered commutative continuous semiring that is proper. If $1 \leq k$ holds for all $k \in K$, $k \neq 0$, then for every alphabet $\Sigma$, $K^{\mathrm{rat}}\langle\!\langle \Sigma^* \rangle\!\rangle$, equipped with the sum order, is a free symmetric inductive $K$-semialgebra generated by $\Sigma$.*

When $K$ satisfies stronger assumptions, then $K^{\mathrm{rat}}\langle\!\langle \Sigma^* \rangle\!\rangle$ is free in the class of inductive $K$-semialgebras.

**Theorem 4.9** *Suppose that $\Sigma$ is an alphabet and $K$ is a commutative inductive semiring that is strongly proper and atomistic. Then $K^{\mathrm{rat}}\langle\!\langle \Sigma^* \rangle\!\rangle$ has the following property: Given any inductive $K$-semialgebra $A$ and function $h : \Sigma \to A$, there is a unique morphism of $K$-semialgebras $h^\sharp : K^{\mathrm{rat}}\langle\!\langle \Sigma^* \rangle\!\rangle \to A$ preserving the star operation which extends $h$.*

1. *If $h^\sharp$ preserves the pointwise ordering on $K^{\mathrm{rat}}\langle\!\langle \Sigma^* \rangle\!\rangle$, for all inductive $K$-semialgebras $A$ and all functions $h : \Sigma \to A$, then, equipped with the pointwise order, $K^{\mathrm{rat}}\langle\!\langle \Sigma^* \rangle\!\rangle$ is a free inductive $K$-semialgebra on $\Sigma$.*

2. *If $K$ is sum ordered and $K^{\mathrm{rat}}\langle\!\langle \Sigma^* \rangle\!\rangle$, equipped with the sum order relation is an inductive semiring and hence an inductive $K$-semialgebra, then $K^{\mathrm{rat}}\langle\!\langle \Sigma^* \rangle\!\rangle$ is a free inductive $K$-semialgebra generated by $\Sigma$.*

*Proof.* The proof is the same as that of Theorem 4.1, but we need a more refined argument to establish that $h^\sharp$ is well-defined. So suppose that $A$ is an inductive $K$-semialgebra and $h : \Sigma \to A$. We define $h^\sharp$ exactly as in the proof of Theorem 4.1. To show that $h^\sharp$ is well-defined, suppose that $\mathcal{A}$ and $\mathcal{B}$ are equivalent $K$-automata over $\Sigma$ in $A$. Then by assumption, $\mathcal{A}$ and $\mathcal{B}$ can be connected by a finite chain of functional, dual functional and invertible diagonal simulations. This yields that $\mathcal{A}h$ and $\mathcal{B}h$ can also be connected by a finite chain of such simulations. Thus, by Corollary 3.10, $\mathcal{A}h$ and $\mathcal{B}h$ are equivalent, proving that $h^\sharp$ is well-defined. □

**Corollary 4.10** *Suppose that $K$ is a commutative idempotent inductive semiring that is strongly proper and atomistic. Then for every alphabet $\Sigma$, $K^{\mathrm{rat}}\langle\!\langle \Sigma^* \rangle\!\rangle$ is a free idempotent inductive $K$-semialgebra generated by $\Sigma$.*

*Proof.* Suppose that $h : \Sigma \to A$ where $A$ is an inductive $K$-semialgebra. Since $K$ is idempotent, the pointwise order of $K^{\mathrm{rat}}\langle\!\langle \Sigma^* \rangle\!\rangle$ agrees with the sum order and is thus preserved by $h^\sharp$. □

**Corollary 4.11** *Suppose that $K$ is a commutative positively ordered locally finite semiring and thus an inductive semiring. Moreover, suppose that $K$ is atomistic and each element of $K$ is a finite sum of elements that have a multiplicative inverse. Then for any alphabet $\Sigma$, $K^{\mathrm{rat}}\langle\!\langle \Sigma^* \rangle\!\rangle$, equipped with the pointwise order, is a free symmetric inductive semiring on $\Sigma$.*

*Proof.* Since $K$ is atomistic, it is equisubstractive [1, 2]. Since $K$ locally finite, it is proper, and since $K$ is equisubstarctive and additively generated by its multiplicative units, it is strongly proper [1]. Since $K$ is locally finite and positively ordered, $K^{\mathrm{rat}}\langle\!\langle \Sigma^* \rangle\!\rangle$ is a (free symmetric) inductive semiring. The result now follows from Theorem 4.9 and Lemma 4.3. (See the proof of Corollary 4.4.) □

**Corollary 4.12** *Suppose that $K$ is a sum ordered commutative continuous semiring that is strongly proper and atomistic. If $1 \leq k$ holds for all $k \in K$, $k \neq 0$, then for every alphabet $\Sigma$, $K^{\mathrm{rat}}\langle\!\langle \Sigma^* \rangle\!\rangle$, equipped with the sum order, is a free inductive $K$-semialgebra generated by $\Sigma$.*

Either of the previous three corollaries may be specialized to give:

**Corollary 4.13** *(Krob [22]) For any alphabet $\Sigma$, $\mathbb{B}^{\mathrm{rat}}\langle\!\langle \Sigma^* \rangle\!\rangle$ is a free idempotent inductive semiring on $\Sigma$.*